\documentclass[12pt]{article}
\pdfoutput=1
\usepackage{epsfig,amsmath,amssymb,bm}
\usepackage{graphicx,psfrag}
\usepackage[top=2cm,bottom=3cm,left=2.4cm,right=2.4cm]{geometry}
\usepackage[titletoc]{appendix}
\usepackage[makeroom]{cancel}
\usepackage{enumitem}
\usepackage{stackrel}
\usepackage{float}
\usepackage[final,color]{showkeys} 
\allowdisplaybreaks

\numberwithin{equation}{section}

\newcommand{\be}{\begin{equation}}
\newcommand{\ee}{\end{equation}}
\newcommand{\bea}{\begin{eqnarray}}
\newcommand{\eea}{\end{eqnarray}}
\newcommand{\bA}{\begin{array}}
\newcommand{\eA}{\end{array}}
\newcommand{\bc}{\begin{center}}
\newcommand{\ec}{\end{center}}

\newcommand{\ra}{\rightarrow}

\newcommand{\ie}{{\it i.e.}}
\newcommand{\eg}{{\it e.g.}}

\newcommand{\scriptA}{\mathcal{A}}
\newcommand{\scriptF}{\mathcal{F}}
\newcommand{\hyperg}{_{2} F_{1}}

\begin{document}

\begin{titlepage}
	
	\bc

	\hfill 
	\\         [22mm]

{\Huge Hyperscaling violation, quasinormal modes\\
          [2mm] and shear diffusion}
    	\vspace{16mm}

Debangshu Mukherjee, K. Narayan\\
\vspace{3mm}
{\small \it Chennai Mathematical Institute, \\ }
{\small \it SIPCOT IT Park, Siruseri 603103, India.\\ }

	\ec
	\medskip
	\vspace{40mm}
	
\begin{abstract}
We study quasinormal modes of shear gravitational perturbations for
hyperscaling violating Lifshitz theories, with Lifshitz and
hyperscaling violating exponents $z$ and $\theta$. The lowest
quasinormal mode frequency yields a shear diffusion constant which is
in agreement with that obtained in previous work by other methods. In
particular for theories with $z< d_i+2-\theta$ where $d_i$ is the
boundary spatial dimension, the shear diffusion constant exhibits
power-law scaling with temperature, while for $z=d_i+2-\theta$, it
exhibits logarithmic scaling. We then calculate certain 2-point
functions of the dual energy-momentum tensor holographically for
$z\leq d_i+2-\theta$, identifying the diffusive poles with the
quasinormal modes above. This reveals universal behaviour
$\eta/s=1/4\pi$ for the viscosity-to-entropy-density ratio for all
$z\leq d_i+2-\theta$.
\end{abstract}

\end{titlepage}

\newpage 
{\footnotesize 
\begin{tableofcontents}
\end{tableofcontents}}

\section{Introduction}

Over the last several years, the framework of gauge/gravity duality
\cite{AdSCFT} has been generalized and applied to understand strongly
coupled non-relativistic field theories: see
\eg\ \cite{Hartnoll:2016apf} for a recent review. A class of these,
dubbed hyperscaling violating Lifshitz (hvLif) theories, has been
studied extensively. These are conformally Lifshitz solutions to
effective Einstein-Maxwell-Dilaton theories. Some of these exhibit
novel entanglement scaling \cite{Ogawa:2011bz,Huijse:2011ef,
Dong:2012se}, reflected also in certain string realizations
\cite{Narayan:2012hk,Singh:2012un,Narayan:2012ks,Narayan:2013qga}.

It is of great interest to understand these nonrelativistic theories
with regard to their low energy behaviour, in particular hydrodynamics
and the viscosity bound \cite{Kovtun:2004de}. Some previous
investigations appear in
\eg\ \cite{Pang:2009wa,Cremonini:2011ej,Hoyos:2013qna,Sadeghi:2014zia,
  Roychowdhury:2014lta,Kiritsis:2015doa,Kuang:2015mlf,Blake:2016wvh,
  Kolekar:2016pnr,Patel:2016ymd,Kolekar:2016yzg}, in part reviewed in
\cite{Hartnoll:2016apf}.  In the context of gauge/gravity duality
\cite{AdSCFT} for relativistic theories, various transport properties
are encoded in the quasinormal modes of the dual gravitational black
branes, see
\eg\ \cite{Kovtun:2005ev,Starinets:2008fb,Nunez:2003eq,Berti:2009kk}.
Quasinormal modes are solutions to the linearized equations governing
the gravitational perturbations that are ingoing at the horizon and
vanishing at the boundary: these boundary conditions make the low
lying hydrodynamic modes damped and diffusive, with a dispersion
relation that encodes the hydrodynamic diffusive poles in
certain 2-point correlation functions in the dual field theory.

Motivated by these earlier studies, in this paper we analyse the
lowest quasinormal mode spectrum for shear gravitational perturbations
in hyperscaling violating Lifshitz theories with Lifshitz exponent $z$
and hyperscaling violating exponent $\theta$. We turn on appropriate
metric and gauge field perturbations $h_{xy}, h_{ty}$ and $a_y$ 
of the form $e^{-i\omega t+iqx}$. Defining appropriate new field variables
${\cal H}$ invariant under a residual gauge symmetry for such
perturbations enables us to identify the relevant differential
equations governing these modes. The hydrodynamic regime allows the
approximation of low frequency and momentum relative to the temperature
scale. Then using $\mathbf{\Omega} \sim {\omega \over T} \ll 1$ and
$\mathbf{Q} \sim {q \over T^{1/z}}\ll 1$, as expansion parameters, we
find series solutions for the quasinormal modes. The lowest quasinormal
modes for these shear perturbations are of the form $\omega=-i{\cal D}q^2$
where ${\cal D}$ is the shear diffusion constant. Our analysis (sec.~2)
of these quasinormal modes and the associated boundary conditions can
be carried out provided the exponents satisfy $z\leq d_i+2-\theta$.
In particular ${\cal D}$ exhibits power law scaling with temperature
for $z< d_i+2-\theta$.
The shear diffusion constant ${\cal D}$ for hvLif theories obtained
thus is in agreement with that obtained previously in
\cite{Kolekar:2016pnr,Kolekar:2016yzg}, where ${\cal D}$ was obtained
by adapting the ``membrane paradigm'' approach of
\cite{Kovtun:2003wp}.  To elaborate further, turning on perturbations
$h_{xy}, h_{ty}$ and $a_y$ and further compactifying the theory along
a spatial direction exhibiting translation invariance, we mapped
near-horizon metric perturbations to gauge field perturbations in an
auxilliary theory in one lower dimension. The gauge fields were used
to define currents $j^{\mu}$ on a ``stretched horizon'', satisfying a
diffusion equation $\partial_t j^t=\mathcal{D}\partial_x^2 j^t$. The
shear diffusion constant $\mathcal{D}$ is obtained by solving for the
perturbations using a set of self-consistent assumptions in a near
horizon expansion. This membrane paradigm approach does not require
holography as such.

In Sec. \ref{correlator} using the asymptotic behaviour of the
quasinormal mode perturbations in sec.~2 above, we adapt the
prescription of \cite{Son:2002sd,Policastro:2002se} to compute certain
2-point correlation functions of the dual energy-momentum tensor
operators.  The poles of these retarded correlators are identical to
the lowest quasinormal frequencies above of the dual black brane for
$z< d_i+2-\theta$, vindicating the correspondence between quasinormal
mode frequencies, the shear diffusion constant and the poles of the
retarded correlators for nonrelativistic theories with $z< d_i+2-\theta$,
thereby giving ${\eta\over s}={1\over 4\pi}$~.

For $z=d_i+2-\theta$, the shear diffusion constant above exhibits
logarithmic scaling, the logarithm containing the ultraviolet cutoff.
However the correlation functions obtained above, in the Kubo limit,
continue to reveal universal behaviour for the viscosity bound with
${\eta\over s}={1\over 4\pi}$ as we discuss in sec.~\ref{sec:log}.
In Appendix A and B, we provide some technical details, and in
Appendix C, we review the membrane paradigm approach to the shear
diffusion constant studied in \cite{Kolekar:2016pnr,Kolekar:2016yzg}.

\section{Hyperscaling violating Lifshitz (hvLif) theory}\label{hvLifqnm}

In this section, we describe the nonrelativistic holographic backgrounds
in which we want to study quasinormal mode solutions. These backgrounds 
of interest here are described by $(d+1)$-dimensional hyperscaling
violating metrics at finite temperature
given by
\be\label{hvmetric}
ds^2 = r^{2\theta/d_i} \left( -\frac{f(r)}{r^{2z}}dt^2 + \frac{dr^2}{f(r)r^2}
+ \sum_{d_i}\frac{dx^2_i}{r^2} \right),
\qquad d_i=d-1 ,\ \ \ f(r)=1-(r_0r)^{d_i+z-\theta} .
\ee
$r=\frac{1}{r_0}$ is the location of the horizon, and $d_i$ is the
boundary spatial dimension. These are conformally Lifshitz solutions to
Einstein-Maxwell-dilaton theories (see Appendix \ref{hvlif-review} for
some details). 
The temperature of the field theory dual to the hvLif theory
(\ref{hvmetric}) is the Hawking temperature of the black brane
\be\label{tempr0}
T = \frac{d_i+z-\theta}{4\pi} r_0^z\ .
\ee

We are interested in studying shear gravitational modes: these are the
modes $h_{xy}$ and $h_{ty}$, which in general couple to the gauge
field perturbations $a_y$. We turn on perturbations of the
form\ $e^{-i\omega t+iqx}h_{\mu \nu}(r) ,\ e^{-i\omega t+iqx}a_{\mu}(r)$,\ and
restrict ourselves to radial gauge ($h_{\mu r}=a_r=0$). Then, as in
\cite{Kolekar:2016pnr,Kolekar:2016yzg}, shear diffusion can be studied
by mapping it to charge diffusion in a theory in one lower dimension
obtained by compactifying one of the spatial directions $x_i$ enjoying
translation invariance, say $y$. Motivated by this, we define the
variables
\be\label{new-var}
H_{ty}=g^{xx}h_{ty}=r^{2-\frac{2\theta}{d_i}}h_{ty}\ , \qquad
H_{xy}=g^{xx}h_{xy}=r^{2-\frac{2\theta}{d_i}}h_{xy}\ .
\ee
Then the equations of motion governing the perturbations are simply
\bea\label{nu-t-eqn}
\partial_r(r^{z+\theta-(d_i+1)}H'_{ty})-ka'_y-\frac{r^{z+\theta-(d_i+1)}}{f}\ q(\omega H_{xy}+q H_{ty})&=&0\ ,\\
\label{nu-x-eqn}
\partial_r(r^{\theta-z-d_i+1}fH'_{xy})+\frac{r^{z+\theta-(d_i+1)}}{f}\ \omega (\omega H_{xy}+q H_{ty})&=&0\ ,\\
\label{nu-r-eqn}
qr^{2-2z}H'_{xy}+\frac{\omega}{f}(H'_{ty}-kr^{(d_i+1)-z-\theta}a_y)&=&0\ ,\\
\label{chi-eqn}
\partial_r(r^{d_i+3-z-\theta}f a'_y)+\frac{r^{d_i+1+z-\theta}}{f}\omega^2 a_y -r^{d_i+3-z-\theta}q^2a_y-kH'_{ty}&=&0\ ,
\eea
where $k=(d_i+z-\theta)\alpha$ and
$\alpha=-\sqrt{\frac{2(z-1)}{d_i+z-\theta}}$\ (see \eqref{bg-gauge-fld}).
The first three equations, namely \eqref{nu-t-eqn},\eqref{nu-x-eqn}
and \eqref{nu-r-eqn} are the three relevant components of the Einstein
equations while \eqref{chi-eqn} is the linearized Maxwell's
equation: we refer to \cite{Kolekar:2016yzg} for details (in part
reviewed in the Appendix). Now, following
\cite{Kovtun:2005ev,Starinets:2008fb}, we note that there is a residual
gauge invariance in these variables representing fluctuations of the
form above: the metric fluctuations transform under infinitesimal
diffeomorphisms as
$h_{\mu\nu}\ra h_{\mu\nu}-\nabla_\mu\xi_\nu-\nabla_\nu\xi_\mu$, with the
gauge functions $\xi_\mu(t,x,r)\equiv \xi_\mu(r) e^{-i\omega t+iq x}$. The
residual gauge invariance then allows us to consider the following gauge
invariant combination
defined as
\begin{equation}
\label{E-defn}
      {\cal H}\ = \omega H_{xy} + q H_{ty}
      - kq \int_{r_c}^r s^{d_i+1-z-\theta}a_y(s) ds\ .
\end{equation}
Note that the gauge field component $a_y$ is invariant by itself.
This combination is motivated by the investigation in \cite{Kolekar:2016yzg}
where a similar combination (\ref{tildeh}) appears as the field variable
(mixing $h_{ty}, a_y$ perturbations) that allows a realization of the
diffusion equation from the Einstein equations governing the near
horizon shear perturbations: this is reviewed in Appendix C. 
Under a compactification of the $y$-direction, these metric components
become gauge field components with a residual $U(1)$ gauge invariance.
We will see the role of this variable ${\cal H}$ in what follows.
The equations of motion \ie\ \eqref{nu-t-eqn}-\eqref{chi-eqn} can be
finally reduced to a system of two coupled second order equations using
the field $\cal{H}$ defined above in \eqref{E-defn}, \ie\ (primes denote
$r$-derivatives)
\bea\label{gen-dimnless-eom}
&& {\cal H''}+ \left[ \partial_r \log r^{z+\theta-(d_i+1)} +  
  \frac{\mathbf{\Omega}^2}{\mathbf{\Omega}^2-(2\pi T)^{2/z-2}\mathbf{Q}^2
    f r^{2-2z}}\partial_r \log(r^{2-2z}f)\right]\cal{H}'\qquad\qquad\qquad
\nonumber\\
&& \qquad\qquad\quad\
+\ (2\pi T)^2\left(\frac{r^{2z-2}}{f^2}\mathbf{\Omega}^2-(2\pi T)^{2/z-2}\frac{\mathbf{Q}^2}{f}\right)\cal{H} \\
&& \qquad\qquad\quad\
+\ (2\pi T)^{2+1/z}k\mathbf{Q}\left(\frac{r^{2z-2}}{f^2}\mathbf{\Omega}^2-(2\pi T)^{2/z-2}\frac{\mathbf{Q}^2}{f}\right)\int_0^r ds\ s^{(d_i+1)-z-\theta}a_y=0\ ,
\qquad\ \nonumber
\eea
\bea \label{chi-full-eqn}
&& a''_y+[\partial_r \log fr^{d_i+3-z-\theta}] a'_y + \left((2\pi T)^2\frac{r^{2z-2}}{f^2}\mathbf{\Omega}^2-(2\pi T)^{2/z}\frac{\mathbf{Q}^2}{f}
-\frac{k^2}{r^2f}\right)a_y \nonumber\\
&& \qquad\qquad\qquad\qquad\qquad\qquad
+\ \frac{(2\pi T)^{1/z-2}k\mathbf{Q}.r^{\theta-z-d_i-1}}{\mathbf{\Omega}^2-(2\pi T)^{2/z-2}\mathbf{Q}^2fr^{2-2z}}{\cal H'}= 0\ .
\eea
Here, $\mathbf{\Omega}$ and $\mathbf{Q}$ satisfying
\begin{equation}\label{dimnless-wq}
\mathbf{\Omega}=\frac{\omega}{2 \pi T}\ ,\ \
\mathbf{Q}=\frac{q}{(2\pi T)^{1/z}}\ ,\qquad
\mathbf{\Omega},\ \mathbf{Q}\ \ll 1\ ,
\end{equation}
are combinations that are dimensionless for the $z=1$ $AdS$ case, and
Lifshitz invariant for $\theta=0$. The condition
$\mathbf{\Omega},\ \mathbf{Q}\ \ll 1$ in (\ref{dimnless-wq}) is
imposed to restrict to the hydrodynamic regime where we can identify
appropriate quasinormal mode solutions and frequencies to the above
equations. These are solutions to the above differential equations
governing the perturbations which are ingoing at the horizon and
vanishing at the boundary (far from the horizon): they are damped
modes reflecting diffusion in these backgrounds and describe how the
perturbed system ``settles down''. In relativistic theories,
quasinormal modes are known to be closely related to hydrodynamic
diffusive modes and the associated diffusive poles in the dual field
theories. In what follows, we will generalize these studies to hvLif
theories.
Restricting to the hydrodynamic regime enables us
to look for solutions to \eqref{gen-dimnless-eom} and \eqref{chi-full-eqn}
in a perturbative series. It turns out that the leading and
next-to-leading behaviour for the mode $\cal H$ can be determined
independent of the $a_y$ solution (which can then be solved for using
the solution of $\cal H$).

\subsection{hvLif in absence of gauge field: dilaton gravity,\ $z=1,\ d_i=2$}

In this subsection, we will analyse a simple case of dilaton gravity
in 4 bulk dimensions ($d_i=2$) as a warmup example. It can be easily
seen from (\ref{hvaction}), that in the absence of a background gauge
field ($A_{\mu}=0$), the hvLif theory reduces to a theory of a scalar
field (dilaton) coupled to gravity with $z=1$ and $k=0$ (which follows
from \ref{bg-gauge-fld}).\  The gauge invariant combination ${\cal H}$
defined in \eqref{E-defn} takes a simpler form and satisfies
\begin{equation}\label{dil-grav-eom}
{\cal H}=\omega H_{xy} + q H_{ty} :\qquad\quad
{\cal H''}- \frac{P'}{P}{\cal H'}+
(2\pi T)^2\left(\frac{\mathbf{\Omega}^2}{f^2}-
\frac{\mathbf{Q}^2}{f}\right){\cal H}=0\ .
\end{equation}
simplifying \eqref{gen-dimnless-eom}, and we have defined the function
$P(r) \equiv \frac{\mathbf{\Omega}^2-\mathbf{Q}^2f(r)}{f(r) r^{\theta-2}}$~.
Close to the horizon (as $r \rightarrow \frac{1}{r_0}$), the above
equation can be approximated as
\begin{equation}
\label{near-hor-dil-g}
      {\cal H''}+\frac{f'}{f}{\cal H'}+\frac{(2\pi T)^2
        \mathbf{\Omega}^2}{f^2}{\cal H}=0\ .
\end{equation}
Using an asymptotic scaling ansatz of the form ${\cal H} \sim f^{A}$
in this equation yields $A=\pm \frac{i\mathbf{\Omega}}{2}$. Choosing
the exponent $A =-\frac{i\mathbf{\Omega}}{2}$, and restoring the
explicit time-dependence, we see that\
${\cal H} \sim e^{-i\omega\left(t+\frac{1}{4\pi T}\log f(r)\right)}$.\
With $z=1, d_i=2$, from (\ref{hvmetric}) we have $f(r)=1-(r_0r)^{3-\theta}$
with the boundary defined at $r\ra 0$. The
blackening factor turns off as $f(r)\ra 1$ far from the horizon only
if $\theta<3$. Focussing therefore on $\theta<3$ from now on, we see
that as time evolves (increasing $t$), these modes carry energy
towards the horizon, \ie\ these are ingoing modes at the horizon.\ 
Taking the ansatz
\be
   {\cal H}(r,\mathbf{\Omega},\mathbf{Q})
   = f(r)^{-\frac{i\mathbf{\Omega}}{2}}F(r,\mathbf{\Omega},\mathbf{Q})\ ,
\ee
and using in \eqref{dil-grav-eom}, we can obtain a second
order equation governing $F(r, \mathbf{\Omega},\mathbf{Q})$. Towards 
studying hydrodynamic modes, we analyse \eqref{dil-grav-eom} in the
regime $\mathbf{\Omega} \ll 1, \mathbf{Q} \ll 1$. To keep track of
the order of the perturbative solution, we introduce a book-keeping
parameter $\lambda$ and rescale $\mathbf{\Omega}\ra \lambda\mathbf{\Omega}$
and $\mathbf{Q}\ra \lambda \mathbf{Q}$, following
\cite{Kovtun:2005ev,Starinets:2008fb}. Rescaling then gives
\begin{equation}
\label{rescaled-eom}
\begin{split}
F''-\left(i\lambda \mathbf{\Omega}\ \frac{f'}{f}+\frac{P'}{P}\right)F'
+\left(-\frac{i\lambda \mathbf{\Omega}}{2}\left(\frac{f'}{f}\right)'-\right.&\left.\frac{\lambda^2\mathbf{\Omega}^2}{4}\left(\frac{f'}{f}\right)^2+
\frac{i\lambda\mathbf{\Omega}}{2}\frac{f'}{f}\frac{P'}{P}\right)F\\
&+(2\pi T)^2\lambda^2\left(\frac{\mathbf{\Omega}^2}{f^2}-\frac{\mathbf{Q}^2}{f}\right)F=0\ .
\end{split}
\end{equation}
Assuming that the solution admits a series expansion in the perturbation
parameter $\lambda$ \ie\
\begin{equation}\label{F(r)exp}
F(r)=F_0(r)+\lambda F_1(r)+\lambda^2 F_2(r)+ \cdots\ ,
\end{equation}
we can write a second order equation for $F_0(r)$ and its corresponding
solution as
\begin{equation}
\label{F0-eqn}
F''_0-\frac{P'}{P}F'_0=0\ ; \qquad
F_0(r)=C_0+C_1\int^r \frac{\mathbf{\Omega}^2-\mathbf{Q}^2f}{fr'^{\theta-2}}\cdot dr'\quad \xrightarrow{\mbox{regularity}}\quad  F_0(r)=C_0\ .
\end{equation}
Near the horizon, $f(r)$ vanishes, giving a logarithmic divergence
in $F_0$. Demanding regularity of the solution at the horizon forces
us to set $C_1=0$, thus giving the solution as simply $F_0=C_0$ in
\eqref{F0-eqn}.
Using this in \eqref{rescaled-eom} and collecting terms of $O(\lambda)$
gives an inhomogeneous differential equation for $F_1(r)$,
\begin{equation}\label{F1-eqn-dg}
  F''_1 -\frac{P'}{P}F'_1=C_0\frac{i\mathbf{\Omega}}{2}\left[
    \left(\frac{f'}{f}\right)'-\frac{P'}{P}\cdot \frac{f'}{f}\right] .
\end{equation}
Integrating and multiplying throughout by $P$, we get\ 
\be\label{kappa1}
F'_1=\frac{i\mathbf{\Omega}}{2}C_0\ \partial_r \log f + \kappa_1 P\quad
\Rightarrow\quad  \kappa_1=-\frac{i C_0}{2\mathbf{\Omega}}f'(1/r_0)r_0^{2-\theta}
=\frac{i C_0}{2\mathbf{\Omega}}(3-\theta)r_0^{3-\theta}\ .
\ee
The above value for the constant $\kappa_1$ is required by demanding
regularity of $F_1$ at the horizon which implies $F'_1$ must be
finite as $r\ra\frac{1}{r_0}$.
Using this value of $\kappa_1$, the solution to \eqref{F1-eqn-dg} is
\begin{equation}\label{dil-g-F1}
  F_1(r)-F_1(1/r_0)
  = \frac{i C_0\mathbf{Q}^2}{2\mathbf{\Omega}}(1-(r_0r)^{3-\theta})\ .
\end{equation}
We set the integration constant $F_1(1/r_0)$ to zero, as in
\cite{Policastro:2002se}. This is consistent with the absence of any 
additional dependence on $\mathbf{\Omega}, \mathbf{Q}$ in the subleading
terms $F_i(r)$ in (\ref{F(r)exp}), \ie\ with fixing the normalization
of the modes as simply $C_0e^{-i\omega t}f(r)^{-i\Omega/2}$ at the horizon.

Imposing the Dirichlet boundary condition ${\cal H}(0)=0$, \ie\ the
fluctuations vanish on the boundary $r=r_c\ra 0$, we obtain
\begin{equation}
  1+\frac{i\mathbf{Q}^2}{2\mathbf{\Omega}}
  \left(1-(r_0r)^{3-\theta}\right)\Big|_{r \sim 0} = 0\ .
\end{equation}
Using (\ref{dimnless-wq}), at the boundary, we thus obtain the dispersion
relation 
\begin{equation}\label{qnmz=1}
\omega = -i \frac{1}{4\pi T}q^2 \equiv -i{\cal D} q^2\ ,
\end{equation}
where $\mathcal{D}=\frac{1}{4\pi T}$ is the shear diffusion constant.
This is consistent with ${\cal D}$ found in
\cite{Kolekar:2016pnr,Kolekar:2016yzg}, using a membrane-paradigm-like
near horizon analysis (generalizing \cite{Kovtun:2003wp}), and the
corresponding guess $\frac{\eta}{s}=\mathcal{D}T=\frac{1}{4\pi}$ for
universal viscosity-to-entropy-density.

\subsection{hvLif theory: generalized analysis}\label{hvLif-section}

In this section, we will study hvLif theories in full generality. To
study hydrodynamics, we will focus on the regime
$\mathbf{\Omega} \ll 1$, $\mathbf{Q}\ll 1$, taking the ansatze
\begin{equation}\label{E-chi-ansatz}
  {\cal H}(r,\mathbf{\Omega},\mathbf{Q})
  =f^{\frac{-i\mathbf{\Omega}}{2}}F(r,\mathbf{\Omega},\mathbf{Q})\ ; \qquad
  a_y(r,\mathbf{\Omega},\mathbf{Q})
  =f^{\frac{-i\mathbf{\Omega}}{2}}G(r,\mathbf{\Omega},\mathbf{Q})\ .
\end{equation}
The factor $f^{-\frac{i\mathbf{\Omega}}{2}}$ reflects the ``ingoing'' nature
of these solutions, as in the previous $z=1$ case. The null energy
condition (\ref{nullee}) implies $d_i+z-\theta >0$ for theories with $z>1$
so the factor $f^{-\frac{i\mathbf{\Omega}}{2}}$ always reflects ``infalling''
modes noting the form of $f(r)$ in (\ref{hvmetric}). Rewriting \eqref{gen-dimnless-eom} in terms of $F$ and $G$ and further rescaling
$\mathbf{\Omega}\ra\lambda\mathbf{\Omega}$,\ $\mathbf{Q}\ra\lambda\mathbf{Q}$, we end up with
\begin{equation}
\label{F-eqn}
\begin{split}
&F''-\left(\frac{H'}{H}+i\lambda \mathbf{\Omega}\frac{f'}{f}\right)F'\\
&-\left[\frac{i\lambda \mathbf{\Omega}}{2}\left(\frac{f'}{f}\right)'+\frac{\lambda^2\mathbf{\Omega}^2}{4}\left(\frac{f'}{f}\right)^2-\frac{i\lambda\mathbf{\Omega}}{2}\frac{f'H'}{fH}-\lambda^2(2\pi T)^2\frac{r^{2z-2}}{f^2}(\mathbf{\Omega}^2-(2\pi T)^{{2\over z}-2}\mathbf{Q}^2fr^{2-2z})\right]F\\
&+\ \lambda^3(2\pi T)^{2+{1\over z}}k\mathbf{Q}\frac{r^{2z-2}}{f^2}(\mathbf{\Omega}^2-(2\pi T)^{{2\over z}-2}\mathbf{Q}^2fr^{2-2z})f^{\frac{i\lambda\mathbf{\Omega}}{2}}\int ds\cdot f^{-\frac{i\lambda\mathbf{\Omega}}{2}}\ s^{d_i+1-z-\theta}G=0\ .
\end{split}
\end{equation}

We also assume the solutions admit a series expansion in $\lambda$ as following
\begin{equation}
\label{pert-ansatz}
\begin{aligned}
F(r,\mathbf{\Omega},\mathbf{Q})&=F_0(r,\mathbf{\Omega},\mathbf{Q})+ \lambda F_1(r,\mathbf{\Omega},\mathbf{Q})+ O(\lambda^2)  + \cdots\\
G(r,\mathbf{\Omega},\mathbf{Q})&=G_0(r,\mathbf{\Omega},\mathbf{Q})+ \lambda G_1(r,\mathbf{\Omega},\mathbf{Q})+ O(\lambda^2) + \cdots
\end{aligned}
\end{equation}
Gathering terms order-by-order, we see that $F_0$ follows a homogeneous second order differential equation while the nature of $F_1$ depends on $F_0$. We have argued in Appendix \ref{F2-details} that the last term in \eqref{F-eqn} becomes relevant only at $O(\lambda^3)$ and does not contribute to the $F_0$ and $F_1$ solutions. In Sec. \ref{ay-solution},  we demonstrate that the $F_0$ and $F_1$ solutions are consistent with \eqref{chi-full-eqn} too and solve for the function $G_0(r,\mathbf{\Omega},\mathbf{Q})$ which is  determined by $F_0$ and $F_1$. Further $G_1$ requires knowledge of $F_2$ also.
Thus although the exact form of the perturbation solutions
${\cal H}, a_y$, is governed by the coupled equations
\eqref{gen-dimnless-eom}, \eqref{chi-full-eqn}, restricting to $O(\lambda)$
essentially decouples the $a_y$ terms from the equation governing
${\cal H}$ which we will solve for below.
 
Sticking \eqref{pert-ansatz} in \eqref{F-eqn} and gathering terms of $O(\lambda^0)$ gives
\begin{equation}
\label{hvlif-F0-eqn}
F''_0-\frac{H'}{H}F'_0=0 \quad \mbox{with} \quad H=\frac{\mathbf{\Omega}^2-(2\pi T)^{2/z-2}\mathbf{Q}^2f r'^{2-2z}}{fr'^{\theta-z-d_i+1}}\quad
\xrightarrow{\mbox{\ regularity\ }}\ \ F_0=C_0\ .
\end{equation} 
To elaborate, the solution to the equation above (analogous to
(\ref{F0-eqn})) is\ $F_0=C_0+C_1\int H dr$ and regularity of $F_0$ 
requires $C_1=0$ giving $F_0=C_0$ above.
The next-to-leading solution $F_1$ satisfies an equation structurally
similar to \eqref{F1-eqn-dg},
\begin{equation}
\label{hvlif-F1-eqn}
F''_1 -\frac{H'}{H}F'_1=C_0\frac{i\mathbf{\Omega}}{2}\left[ \left(\frac{f'}{f}\right)'-\frac{H'}{H}\cdot \frac{f'}{f}\right]\ .
\end{equation}
Integrating gives
\begin{equation}
\label{hvLif-F1-eqn}
F'_1=\frac{i\mathbf{\Omega}C_0}{2}\partial_r \log f+\kappa_2\cdot
\frac{\mathbf{\Omega}^2-(2\pi T)^{2/z-2}\mathbf{Q}^2f r^{2-2z}}
     {fr^{\theta-z-d_i+1}} \quad\Rightarrow\quad
\kappa_2=\frac{iC_0}{2\mathbf{\Omega}}(d_i+z-\theta)r_0^{d_i+z-\theta}\ .
\end{equation}
The integration constant $\kappa_2$ is fixed as in (\ref{kappa1}) by
demanding regularity of $F_1$ at the horizon $r\ra 1/r_0$. This forces
the singular part from the first term to be cancelled by the other
singular piece coming from the $O(\mathbf{\Omega}^2)$ term, fixing
$\kappa_2$ above. 
The value of $\kappa_2$ can be used to write down the solution to
$F_1(r)$,
\be\label{gen-F1}
F_1(r)  
= -\frac{iC_0(d_i+z-\theta)\mathbf{Q}^2}{2\mathbf{\Omega}}(2\pi T)^{\frac{2}{z}-2}r_0^{d_i+z-\theta}\int_{\frac{1}{r_0}}^r r'^{d_i+1-z-\theta}dr'\ .
\ee
As in (\ref{dil-g-F1}) and the comments following it, we have set the
integration constant $F_1(1/r_0)$ to zero in the second line.
Then the solution to \eqref{gen-dimnless-eom} upto first order in the
hydrodynamic expansion can be written down and varies depending on the
value of $(d_i,z,\theta)$.

\vspace{2mm}

\noindent \underline{{\bf $\bm{z< d_i+2-\theta}$}}:\ \ This is the sector
continuously connected to relativistic ($AdS$) theories which have
$z=1, \theta=0$.\ This sector also includes hvLif theories arising
from reductions of $p\leq 4$ nonconformal $Dp$-branes where\
$z=1,\ d_i=p,\ \theta=p-{9-p\over 5-p}$.\
The solution to \eqref{gen-dimnless-eom} upto first order is given by
\be\label{full-E-soln}
{\cal H} = C_0f(r)^{-\frac{i\mathbf{\Omega}}{2}} \left[ 1 +
  \frac{iq^2}{(d_i+2-z-\theta)\omega}r_0^{z-2}\cdot (1-(r_0r)^{d_i+2-z-\theta})
  \right]\ ,
\ee
where $r_0$ is related to the temperature $T$ as in (\ref{tempr0}).
Imposing Dirichlet boundary conditions \ie\ ${\cal H} (r \ra 0)=0$ at
the UV cut-off boundary ($r=r_c\ra 0$) using \eqref{dimnless-wq} gives
\begin{equation}
\label{dispersion-reln}
\omega = -iq^2\cdot \frac{1}{d_i+2-z-\theta}\cdot
\left(\frac{4\pi T}{d_i+z-\theta}\right)^{1-2/z}\ \equiv\
-i{\cal D}q^2\ ,
\end{equation}
as the quasinormal mode frequency. This gives the leading shear diffusion
constant\
\be\label{genDiffconst}
\mathcal{D}\ =\frac{r_0^{z-2}}{d_i+2-z-\theta}\ =\
\frac{1}{d_i+2-z-\theta}\cdot
\Big(\frac{4\pi }{d_i+z-\theta}\Big)^{1-2/z}T^{\frac{z-2}{z}}\ ,
\ee
which matches the result obtained using the membrane paradigm approach
in \cite{Kolekar:2016pnr,Kolekar:2016yzg}\ (reviewed in Appendix C).
This led to a guess for the relation between the shear diffusion constant
$\mathcal{D}$ and shear viscosity $\eta$, consistent with various special
cases,
\begin{equation}
\label{eta-s-D-T}
\frac{\eta}{s}=\frac{d_i+2-z-\theta}{4\pi}\left(\frac{4\pi}{d_i+z-\theta}\right)^{\frac{2-z}{z}}\mathcal{D}T^{\frac{2-z}{z}}=\frac{1}{4\pi}\ .
\end{equation}
We will later (sec.~3) evaluate the viscosity using holographic
techniques, corroborating this. For $z=1$,
we see that $\frac{\eta}{s}={\cal D}T={1\over 4\pi}$ as in (\ref{qnmz=1}).
It is worth noting that these quasinormal modes are diffusive damped
modes. Our analysis and results hold in the regime (\ref{dimnless-wq})
so in particular ${q\over T^{1/z}}\ll 1$. The quasinormal mode frequency
(\ref{dispersion-reln}) can then be expressed as\
$\omega \sim -i ({q\over T^{1/z}})^2 T$,\ 
and we are working at finite temperature in the hydrodynamic low
frequency, low momentum regime. Then the time dependence of these modes
is $\sim e^{-i\omega t}\sim e^{-\Gamma t}$, damped on long timescales.

\vspace{1mm}

\noindent \underline{{\bf $\bm{z=d_i+2-\theta}$}}:\ \ Here, the integral
in \eqref{gen-F1} gives the solution to \eqref{gen-dimnless-eom} to first
order as
\be\label{qnmHlogcase}
{\cal H}(r)
= C_0f(r)^{-\frac{i\mathbf{\Omega}}{2}}\left[1+\frac{iq^2}{\omega}r_0^{z-2}\log \frac{1}{r_0r}\right]\ ,
\ee
where (\ref{tempr0}) now gives $T={z-1\over 2\pi} r_0^z$.\
Then defining\ $\Lambda = \frac{z-1}{2\pi}\frac{1}{r_c^z}$\ gives the
low-lying quasinormal frequency
\begin{equation}
\label{logdiff}
  \omega =-iq^2\cdot \frac{1}{z}\left(\frac{2\pi}{z-1}\right)^{1-2/z}\cdot
  T^{\frac{z-2}{z}}\log \frac{\Lambda}{T}\ \equiv -i{\cal D}q^2\ .
\end{equation}
This gives the shear diffusion constant\
$\mathcal{D}=\frac{1}{z}\left(\frac{2\pi}{z-1}\right)^{1-2/z}\cdot T^{\frac{z-2}{z}}\log \frac{\Lambda}{T}$
scaling logarithmically with temperature alongwith a power-law pre-factor.
This also agrees with the results in \cite{Kolekar:2016pnr,Kolekar:2016yzg}.
The logarithmic scaling necessitating the ultraviolet scale
$\Lambda$ perhaps suggests that this leading relation for the
quasinormal mode frequency is subject to subleading corrections and
possibly appropriate resummations. Nevertheless, recasting as\
$\omega\sim -i ({q\over T^{1/z}})^2 T\log \frac{\Lambda}{T}$ shows that
in the hydrodynamic regime ${q\over T^{1/z}}\ll 1$, this leading mode is 
diffusive with damped time-dependence: in fact for $T\ll \Lambda$, the
extra $log$-factor leads to additional damping. The hvLif theories
arising from null reductions of $AdS$ and nonconformal brane plane waves
\cite{Narayan:2012hk,Singh:2012un,Narayan:2013qga} have exponents
satisfying $z=d_i+2-\theta$: taking the quasinormal modes as a measure
of stability of the backgrounds, we see that the diffusive frequencies
suggest that low lying modes do not indicate any instability. The
logarithmic behaviour of the leading shear diffusion constant then
suggests a possibly novel limit of hydrodynamics in these theories,
perhaps stemming from the large boost in the above string constructions.

\vspace{1mm}

\noindent \underline{{\bf $\bm{z>d_i+2-\theta}$}}:\ \ The integral in
\eqref{gen-F1} scales as $r_c^{d_i+2-\theta-z}$ thus acquiring
dominant (divergent) contribution from high energy scales near
$r_c\sim 0$. There is no universal low energy behaviour emerging
from near horizon physics: it appears that these methods fail to yield
insight on quasinormal modes, as does the membrane paradigm approach
\cite{Kolekar:2016pnr,Kolekar:2016yzg}. This sector includes \eg\
reductions of $D6$-branes ($d_i=6,\ z=1,\ \theta=9$) with ill-defined asymptotics.

\subsubsection{Solving for gauge field perturbation $a_y$}
\label{ay-solution}
Using the ansatze \eqref{E-chi-ansatz} and rescaling $\mathbf{\Omega}\rightarrow \lambda \mathbf{\Omega}$ and $\mathbf{Q} \rightarrow \lambda \mathbf{Q}$ we can recast \eqref{chi-full-eqn} as
\begin{equation}
\label{G-full-eqn}
\begin{split}
G''&-\frac{i\lambda \mathbf{\Omega}}{2}\cdot \frac{f'}{f}G'+\partial_r \ln fr^{d_i+3-z-\theta}G'-\frac{i\lambda \mathbf{\Omega}}{2}\left(\frac{f'}{f}\right)'G-\frac{\lambda^2\mathbf{\Omega}^2}{4}\left(\frac{f'}{f}\right)^2G\\
&-\frac{i\lambda \mathbf{\Omega}}{2}\frac{f'}{f}\partial_r \ln fr^{d_i+3-z-\theta}\ G+\lambda^2(2\pi T)^2\frac{r^{2z-2}}{f^2}(\mathbf{\Omega}^2-(2\pi T)^{2/z-2}\mathbf{Q}^2fr^{2-2z})\ G\\
&-\frac{k^2}{r^2f}\ G +\frac{(2\pi T)^{1/z-2}k\mathbf{Q}.r^{\theta-z-d_i-1}}{\mathbf{\Omega}^2-(2\pi T)^{2/z-2}\mathbf{Q}^2fr^{2-2z}}\cdot\frac{1}{\lambda}\left(F'-\frac{i\lambda \mathbf{\Omega}}{2}\frac{f'}{f}F\right)=0\ .
\end{split}
\end{equation}
Plugging in the series ansatz \eqref{pert-ansatz} we can construct the
perturbative solution for $a_y$ order-by-order. The leading order
equation appears at $O({1 \over \lambda})$ and is given by $F'_0=0$
giving $F_0=const$: this can be seen to be consistent with
\eqref{hvlif-F0-eqn}. We will subsequently see that $G_0$ is
determined by $F_1$ and $F_0$ while $G_1$ is determined by $F_2$ and
$G_0$, and so on. More generally, all subsequent equations involve
more variables so there is no inconsistency in the solutions due to
potential overconstraining in this system of equations.

Gathering all terms of $O(\lambda^0)$, we see $G_0$ follows the equation:
\begin{equation}
\label{G0-r-eqn}
G''_{0}+\partial_r \log fr^{d_i+3-z-\theta}G'_{0}-\frac{k^2}{r^2f}G_0+ \frac{(2\pi T)^{1/z-2}k\mathbf{Q}.r^{\theta-z-d_i-1}}{\mathbf{\Omega}^2-(2\pi T)^{2/z-2}\mathbf{Q}^2fr^{2-2z}}\left(F'_1-\frac{i\mathbf{\Omega}}{2} \frac{f'}{f} F_0\right)=0\ .
\end{equation}
For $z<d_i+2-\theta$, using \eqref{hvlif-F0-eqn} and \eqref{gen-F1} we can write the most general solution to the above equation in terms of a new radial variable as
\begin{equation}
\label{G0-soln}
G_0(x)=\frac{A}{mn}+C_1x^n\ \hyperg \left[2, \frac{n}{m},2+\frac{n}{m}; x^m\right]-\frac{C_2}{n}x^{-m}(n+(m-n)x^m)\ ,
\end{equation}
where 
\begin{equation}
\label{xmnA-defn}
x=r_0r\ , \quad m=d_i+z-\theta\ , \quad n=2z-2\ , \quad
A=iC_0kr_0^{d_i-\theta}\frac{q}{\omega}\ ,
\end{equation}
while $C_1$ and $C_2$ are arbitrary constants which are to be fixed by
demanding regularity of $a_y$ at the horizon. Potential divergences in
$a_y(r)$ and $a'_y(r)$ near the horizon can be removed by choosing
$C_1=0$ and $C_2=\frac{A}{m^2}$. In terms of the original radial
coordinate $r$, the solution is
\begin{equation}
\label{G0-r-soln}
G_0(r)=-iC_0k\ \frac{q}{\omega}\cdot\frac{r_0^{d_i-\theta}}{(d_i+z-\theta)^2}\cdot (r_0r)^{-(d_i+z-\theta)}f(r)\ .
\end{equation}
For relativistic theories, ($z=1, k=0$) the above expression vanishes identically implying that the shear mode sector is governed exclusively by metric perturbations $H_{xy}$ and $H_{ty}$.

The subleading term in $a_y$ \ie\ $G_1(r)$ can be determined by collecting terms of $O(\lambda)$ from \eqref{G-full-eqn}. The inhomogeneous part of the equation governing $G_1$ involves $F'_2$ and $F_1$. $F'_2(r)$ can be evaluated from $O(\lambda^2)$ terms of \eqref{F-eqn}. Although we could find the general solution to $G_1$, finding the integration constants respecting regularity at the horizon seems difficult and cumbersome by analytic means. We discuss further details about $F_2$ and $G_1$ in Appendix \ref{F2-details}.

\section{Dual field theory correlation functions}
\label{correlator}

In this section, we will determine the energy-momentum tensor correlation
functions $\langle TT\rangle$ following the prescription in
\cite{Son:2002sd,Policastro:2002se}, and defining $T_{\mu\nu}$ as dual to
the perturbation $h_{\mu\nu}$.
The action governing the perturbations using the variables $H_{ty},
H_{xy}$ and $a_y$ in \eqref{new-var} is given by
\begin{equation}
\begin{aligned}
S^{pert} 
=&-\frac{1}{16\pi G_N^{(d_i+2)}}\int d^{d_i+2}x\ \left[\frac{1}{2}r^{\theta-z-d_i+1}\Big(-r^{2z-2}(H'_{ty})^2+\frac{r^{2z-2}}{f}(\omega H_{xy}+q H_{ty})^2\right.\\
&\left.+\ f(H'_{xy})^2\Big)-kH_{ty} a'_y+\frac{1}{2} r^{d_i+3-z-\theta}f(a'_y)^2+\frac{1}{2} r^{d_i+3-z-\theta}\left(\frac{r^{2z-2}}{f}\omega^2-q^2\right)a_y^2\right] .
\end{aligned}
\end{equation}
The equations of motion from this action lead to
(\ref{nu-t-eqn})-(\ref{chi-eqn}) and we have suppressed contact terms.
The above expression can be recast as a bulk piece (which vanishes by the
equations of motion) and a boundary term. This boundary action takes the
form 
\begin{equation}
S^{bdy}=-\frac{1}{32 \pi G_{N}^{(d_i+2)}}\int d^{d_i+1}x\ \Big[r^{\theta-z-d_i+1} \left(-r^{2z-2}H_{ty}H'_{ty} +f H_{xy}H'_{xy}\right)+r^{d_i+3-z-\theta}a_ya'_y\Big]+\cdots
\end{equation}
again suppressing contact terms.
Using the equation of motion \eqref{nu-r-eqn} and the definition of
$\cal H$ \eqref{E-defn}, we can recast the relevant terms of the above
action as
\begin{equation}
\begin{split}
  S^{bdy}=\lim_{r \to r_c}-\frac{1}{32 \pi G_N^{(d_i+2)}}\int d^{d_i+1}x\ &
  \left[\frac{fr^{\theta-z-d_i+1}}{\omega^2-q^2fr^{2-2z}}\ {\cal H'}(r,x)
    {\cal H}(r,x)\right]+\cdots 
    \end{split}
\end{equation}
In the above equation, we have exhibited only those terms that contribute
to the 2-point function \ie\ terms that are at least second order in
$\cal H$. 
Using the Fourier decomposition of the bulk field as
${\cal H}(r,t,x)=\int d\omega\ dq\ e^{-i\omega t+iqx}{\cal H}(r,\omega,q)$
we can further recast as
\begin{equation}
\label{bdy-action}
S^{bdy} =\lim_{r \to r_c}\ -\frac{1}{32 \pi G_N^{(d_i+2)}}\int  d\omega\ dq\
\frac{fr^{\theta-z-d_i+1}}{\omega^2-q^2fr^{2-2z}}\
     {\cal H'}(r,\omega,q){\cal H}(r,-\omega,-q)\ .
\end{equation}
For $z=1, \theta=0$, this agrees with the $AdS$ case in
\cite{Kovtun:2005ev}. Also, for theories with Lifshitz symmetry, it is
clear that (\ref{bdy-action}) is Lifshitz-invariant. The
$\langle TT \rangle$ shear correlator in the boundary theory is defined as
\begin{equation}
\label{correlator-defn}
G_{xy,xy}=\langle T_{xy}T_{xy} \rangle = \left.\frac{\delta^2 S^{bdy}}{\delta h^{x(0)}_y \delta h^{x(0)}_y}\right|_{r \approx r_c}\ .
\end{equation}
We define the boundary fields through the $r\ra r_c\sim 0$ limits of the
bulk fields, \ie\ ${\cal H}^{(0)}(\omega,q)={\cal H}(r_c,\omega,q)$ and
$h^{x(0)}_y(\omega,q)=h^{x(0)}_y(r_c,\omega,q)$.
The asymptotics can be analysed by studying \eqref{gen-dimnless-eom} in
the limit $r \rightarrow 0$ at zero momenta and frequency
($\mathbf{Q}=\mathbf{\Omega} =0$) \ie\
\begin{equation}\label{HasymptEqn}
{\cal H''}+\frac{z+\theta-d_i-1}{r}{\cal H'}=0\ .
\end{equation}
The solutions are ${\cal H} =Ar^{\Delta}$\ with\ 
$\Delta =0,\ d_i+2-z-\theta$.\ \ Thus in the hydrodynamic regime \ie\
$\mathbf{\Omega, Q} \ll 1$, we can schematically write the mode as
${\cal H}= \mathcal{A}(\omega,q) + \mathcal{B}(\omega,q)r^{d_i+2-z-\theta}$
where the functions $\mathcal{A}(\omega,q)$ and $\mathcal{B}(\omega,q)$
can be read off from \eqref{full-E-soln},
\begin{equation}\label{asymptFalloffs}
\mathcal{A}(\omega,q)= C_0\left[1+ i \frac{q^2}{\omega}\frac{r_0^{z-2}}{d_i+2-z-\theta} \right]\ , \qquad \quad
\mathcal{B}(\omega,q)=-C_0\frac{iq^2}{\omega}\frac{r_0^{d_i-\theta}}{d_i+2-z-\theta}\ .
\end{equation}
We can write the normalized bulk field ${\cal H}$ in terms of its source
${\cal H}^0$
\begin{equation}\label{normalizedH}
 {\cal H}(\omega, q)={\cal H}^{(0)}(\omega,q)\frac{1}{{\cal N}}\left[1 +
 \frac{\mathcal{B}(\omega,q)}{\mathcal{A}(\omega,q)}r^{d_i+2-z-\theta}\right]\ ,
    \qquad\quad {\cal N} = 1+ \frac{\mathcal{B}(\omega,q)}{\mathcal{A}(\omega,q)}r_c^{d_i+2-z-\theta}\ .
\end{equation}
Note that the normalization factor ${\cal N}$ satisfies ${\cal N}\sim 1$
as $r_c\ra 0$ for $z<2+d_i-\theta$.\ 
Using this solution in \eqref{bdy-action}, we get
\begin{equation}
S^{bdy}=\lim_{r \ra r_c} -\int \frac{d\omega\ dq} {32 \pi G_N^{(d_i+2)}}\
\frac{fr^{2-2z}}{\omega^2-q^2fr^{2-2z}}\cdot \frac{d_i+2-z-\theta}{{\cal N}}
\cdot\frac{\mathcal{B}(\omega,q)}{\mathcal{A}(\omega,q)} 
\cdot {\cal H}^{(0)}(\omega,q){\cal H}^{(0)}(-\omega,-q)
\end{equation}
As $r\ra r_c\ra 0$ with $z>1$, we note that\
$\lim_{r \ra 0}\frac{fr^{2-2z}}{\omega^2-q^2fr^{2-2z}} \frac{1}{{\cal N}^2}
\frac{\mathcal{B}(\omega,q)}{\mathcal{A}(\omega,q)}
= -\frac{1}{q^2}\frac{\mathcal{B}(\omega,q)}{\mathcal{A}(\omega,q)}$~.\ 
From the definition of $\cal H$ in \eqref{E-defn} and also noting
$H_{xy}\equiv h^x_y=h^y_x$ and $H_{ty}\equiv h^y_t$ from \eqref{new-var},
we see that\
$\frac{\delta}{\delta H^{(0)}_{xy}}=\frac{\delta}{\delta h^{x(0)}_{y}}=\omega \frac{\delta}{\delta {\cal H}^{(0)}}$.\ 
Thus the correlation function \eqref{correlator-defn} becomes
\begin{equation}
\begin{aligned}
\label{xy-xy-corr}
G_{xy,xy}=\langle T^x_y(k)T^x_y(-k) \rangle=\frac{\delta^2 S^{bdy}}{\delta h^{x(0)}_y(k) \delta h^{x(0)}_y(-k)}&= \omega^2\frac{\delta^2 S^{bdy}}{\delta {\cal H}^{(0)}(k) \delta {\cal H}^{(0)}(-k)}\\
&=\frac{1}{16\pi G_N^{(d_i+2)}}\frac{i\omega^2 r_0^{d_i-\theta}}{\omega+i\mathcal{D}q^2}\ ,
\end{aligned}
\end{equation}
with $\mathcal{D}$ given in ({\ref{genDiffconst}), and $k=(\omega,q)$.
The Kubo formula then gives viscosity as
\begin{equation}
\label{kubo}
\eta = \lim_{\omega \to 0}\frac{G_{xy,xy}(\omega,q=0)}{i\omega}
= \frac{r_0^{d_i-\theta}}{16\pi G_N^{(d_i+2)}}\ .
\end{equation}
With the entropy density given in terms of the horizon area
$s = \frac{r_0^{d_i-\theta}}{4G_N^{(d_i+2)}}$, we obtain universal behaviour
for the viscosity bound $\frac{\eta}{s}=\frac{1}{4\pi}$, as for 
relativistic theories \cite{Kovtun:2004de}. This is consistent
with \cite{Kolekar:2016pnr,Kolekar:2016yzg}, where we conjectured the universal relation \eqref{eta-s-D-T} saturating the proposed viscosity bound in \cite{Kovtun:2004de}.
Also, we can write down other correlators as follows:
\begin{equation}
\label{yt-corr}
\begin{aligned}
G_{ty,ty}=\langle T^{y}_t(k) T^y_t(-k)\rangle&=\frac{1}{16\pi G_N^{(d_i+2)}}\frac{iq^2r_0^{d_i-\theta}}{\omega+i\mathcal{D}q^2}\ ,\\
G_{ty,xy}=\langle T^y_t(k)T^y_x(-k) \rangle&=\frac{1}{16\pi G_N^{(d_i+2)}}\frac{i\omega q r_0^{d_i-\theta}}{\omega+i\mathcal{D}q^2}\ .
\end{aligned}
\end{equation} 
Each correlator above exhibits a pole at $\omega =-i{\cal D}q^2$
which is the lowest lying quasinormal mode as we have seen earlier.
The viscosity (\ref{kubo}) above agrees with the result in
\cite{Kuang:2015mlf}: however what is noteworthy in our analysis is
that this is obtained in the regime $z<d_i+2-\theta$.

\subsection{Dual field theory correlation functions: $z=d_i+2-\theta$}
\label{sec:log}

hvLif theories with $z=d_i+2-\theta$ arise from the null reductions of
highly boosted black branes in \cite{Narayan:2012hk,Narayan:2013qga} as
mentioned previously. The asymptotic fall-offs in (\ref{HasymptEqn}),
(\ref{asymptFalloffs}), coincide in this case: this is the interface of
the standard/alternative quantization in \cite{Klebanov:1999tb}, and
one of the solutions contains a logarithm.
We see that in this case, \eqref{HasymptEqn} reduces to 
\begin{equation}
{\cal H}''+\frac{1}{r}{\cal H}'=0\ ,
\end{equation} 
with the solution ${\cal H}= \mathcal{A}(\omega,q)+\mathcal{B}(\omega,q)\log\frac{1}{r_0r}$ in (\ref{qnmHlogcase}). We define the normalized bulk field
in terms of the source as
\begin{equation}\label{normalHlog}
{\cal H}(r,\omega,q)={\cal H}_0(\omega,q) \cdot \frac{1+\frac{iq^2}{\omega}r_0^{z-2}\log \frac{1}{r_0r}}{1+\frac{iq^2}{\omega}r_0^{z-2}\log \frac{1}{r_0r_c}}\ ,
\end{equation}
the source being the boundary value
${\cal H}^{(0)}(\omega, q)={\cal H}(r_c,\omega,q)$.\ Note that unlike
${\cal N}$ in (\ref{normalizedH}), the logarithm does not die near the
boundary and the normalization above is less trivial.
Using (\ref{normalHlog}), the relevant part of the boundary action becomes
\begin{equation}
\begin{aligned}
S^{bdy}&=\lim_{r \rightarrow r_c}-\frac{1}{32 \pi G_N^{(d_i+2)}}\int d^{d_i+1}x
\left[\frac{fr^{3-2z}}{\omega^2-q^2fr^{2-2z}}\ {\cal H'}(r,x)
{\cal H}(r,x)\right]  \\ 
&= \frac{1}{32 \pi G_N^{(d_i+2)}}\int d\omega. dq\ \frac{ir_0^{z-2}}{\omega}\frac{1}{\left(1+\frac{iq^2}{\omega}r_0^{z-2}\log\frac{1}{r_0r_c}\right)}{\cal H}^{(0)}(\omega,q){\cal H}^{(0)}(-\omega,-q)\ .
\end{aligned}
\end{equation}
As before, we obtain the energy-momentum tensor correlation functions as
\begin{equation}
\begin{aligned}
G_{xy,xy}=\langle T^x_y(k)T^x_y(-k) \rangle=\frac{\delta^2 S^{bdy}}{\delta h^{x(0)}_y(k) \delta h^{x(0)}_y(-k)}&= \omega^2\frac{\delta^2 S^{bdy}}{\delta {\cal H}^{(0)}(k) \delta {\cal H}^{(0)}(-k)}\\
&=\frac{1}{16\pi G_N^{d_i+2}}\frac{i\omega^2 r_0^{z-2}}{\omega+i\mathcal{D}q^2}\ ,
\end{aligned}
\end{equation}
using the expression for the shear diffusion constant in \eqref{logdiff}.
Using the Kubo formula, we again see that 
\begin{equation}
\label{eta-kubo}
\eta = \lim_{\omega \to 0}\frac{G_{xy,xy}(\omega,q=0)}{i\omega}
= \frac{r_0^{z-2}}{16\pi G_N^{(d_i+2)}}
= \frac{r_0^{d_i-\theta}}{16\pi G_N^{(d_i+2)}}\  .
\end{equation}
Note that the diffusive pole again coincides with the quasinormal mode
frequency in (\ref{logdiff}): note that this here arises from the
nontrivial normalization factor in (\ref{normalHlog}). Now using the
entropy density $s = \frac{r_0^{d_i-\theta}}{4G_N^{(d_i+2)}}$ given in
terms of the horizon area, we again find universal behaviour
${\eta\over s}=\frac{1}{4\pi}$ for the viscosity bound, again as for
relativistic theories \cite{Kovtun:2004de}. It is worth noting that in
applying the Kubo formula, we first restrict to the zero momentum
sector $q=0$, which kills off the term containing the leading diffusion
constant ${\cal D}$ which, strictly speaking, is logarithmically
divergent as $r_c\ra 0$.

\section{Discussion}\label{discussions}

We have studied low lying hydrodynamic quasinormal modes for shear
perturbations of hyperscaling violating Lifshitz black branes: these
are of the form $\omega=-i{\cal D}q^2$ where ${\cal D}$ is the shear
diffusion constant. This is consistent with ${\cal D}$ obtained in
\cite{Kolekar:2016pnr,Kolekar:2016yzg} through a membrane-paradigm
analysis of near horizon perturbations and the associated shear
diffusion equation. This shear diffusion equation $\partial_t
j^t=\mathcal{D}\partial_x^2 j^t$ following from the second order
Einstein equations in a sense dictates the form of ${\cal D}$ above
and is consistent with the low-lying quaisnormal mode spectrum. The
analysis here and the associated boundary conditions are valid for
theories with exponents satisfying $z\leq d_i+2-\theta$: this is the
regime that is continuously connected to $AdS$ theories ($z=1,
\theta=0$). Using the asymptotics of these quasinormal modes, retarded
correlators of dual operators can be obtained: we have seen that the
poles of the retarded $\langle TT\rangle$ correlator at finite
temperature coincide with the lowest quasinormal frequencies of the
dual gravity theory. This analysis appears consistent with the Kubo
formula for viscosity via the retarded Green's function at zero
momentum only for theories with $z\leq d_i+2-\theta$. Perhaps this is
not surprising given the asymptotic fall-offs of the quasinormal
modes: for $z> d_i+2-\theta$, high energy modes appear to dominate,
with no universal low energy behaviour for the diffusion expression.
It would be interesting to understand this better.

hvLif spacetimes with $z=d_i+2-\theta$ exhibit more interesting
hydrodynamic behaviour as we have seen: the asymptotic fall-offs of
the bulk modes coincide here. While correlation functions in the Kubo
limit continue to reveal universal behaviour ${\eta\over s}={1\over
  4\pi}$ for the viscosity, the leading shear diffusion constant
exhibits logarithmic scaling (involving the ultraviolet cutoff),
perhaps suggesting that subleading contributions are important with
some resummation required. The null reductions
\cite{Narayan:2012hk,Narayan:2013qga} of highly boosted black branes
give hvLif theories with $z=d_i+2-\theta$. It is worth noting that the
boost induces anisotropy in the system\ (although the lower dimensional
theory after compactification enjoys translation invariance in the
spatial directions so that the zero momentum Kubo limit studied here
is unambiguous). The large boost involved in these string constructions
possibly leads to novel hydrodynamic behaviour.  It would perhaps be
interesting to study this directly from the null reduction of black
brane hydrodynamics and the fluid/gravity correspondence
\cite{Bhattacharyya:2008jc}.

\vspace{7mm}

{\footnotesize \noindent {\bf Acknowledgements:}\ \ It is a pleasure
to thank Kedar Kolekar, Alok Laddha, R. Loganayagam and Mukund
Rangamani for helpful discussions. DM thanks the organizers of the
ICTP Spring School on Superstring Theory, Trieste, Italy, and the
Asian Winter School on Strings, Particles and Cosmology, Sun Yat-sen
University, China, and the String Theory Groups at TIFR Mumbai, HRI
Allahabad, and ICTS Bangalore, for hospitality while this work was in
progress.​ This work is partially supported by a grant to CMI from the
Infosys Foundation.}

\appendix
\section{Reviewing hyperscaling violating Lifshitz spacetimes}
\label{hvlif-review}

The metric \eqref{hvmetric} is a solution to the Einstein-Maxwell-Dilaton
action
\begin{equation}
\label{hvaction}
S= -\frac{1}{16 \pi G_{N}^{(d+1)}}\int d^{d+1}x\  \sqrt{-G}\left[R-\frac{1}{2}\partial_{\mu}\phi\partial^{\mu}\phi-\frac{Z(\phi)}{4}F_{\mu \nu}F^{\mu \nu}+V(\phi) \right]\ ,
\end{equation}
where the various fields and parameters appearing in the action are listed as follows:
\begin{eqnarray}
\phi&=&\sqrt{2(d_i-\theta)(z-\theta/d_i-1)}\ \log r\ ,\\
\label{bg-gauge-fld}
A_t&=&\frac{\alpha f(r)}{r^{d_i+z-\theta}}\ , \qquad
\alpha=-\sqrt{\frac{2(z-1)}{d_i+z-\theta}}\ , \qquad A_i=0\ .\\
V(\phi)&=&(d_i+z-\theta)(d_i+z-\theta -1)r^{-\frac{2\theta}{d_i}}\ ; \qquad 
Z(\phi)=r^{\frac{2\theta}{d_i}+2d_i-2\theta}=e^{\lambda \phi}\ .
\end{eqnarray}
The null energy conditions following from (\ref{hvmetric}) give constraints
on the Lifshitz $z$ and hyperscaling violating $\theta$ exponents\ 
\be\label{nullee}
(z-1)(d_i+z-\theta)\geq 0\ ,\qquad (d_i-\theta)(d_i(z-1)-\theta)\geq 0\ .
\ee
Varying with $G_{\mu \nu}$, $A_{\mu}$ and $\phi$, we obtain the following
equations of motion,
\begin{equation}\label{einsteineqn}
R_{\mu\nu}=\frac{1}{2}\partial_{\mu}\phi\partial_{\nu}\phi -G_{\mu\nu}\frac{V(\phi)}{d-1} + \frac{Z(\phi)}{2}G^{\rho\sigma}F_{\rho\mu}F_{\sigma\nu}
- \frac{Z(\phi)}{4(d-1)}G_{\mu\nu}F_{\rho\sigma}F^{\rho\sigma}\ ,
\end{equation}
\begin{equation}\label{gaugeeqn}
\nabla_{\mu}(Z(\phi)F^{\mu\nu})=0\ , \qquad \frac{1}{\sqrt{-G}}\partial_{\mu}(\sqrt{-G}G^{\mu\nu}\partial_{\nu}\phi)+\frac{\partial V(\phi)}{\partial\phi}-\frac{1}{4}\frac{\partial Z(\phi)}{\partial\phi}F_{\rho\sigma}F^{\rho\sigma}=0\ .
\end{equation}
Turning on gravitational, gauge field and scalar field
perturbations $h_{\mu \nu}(x)$, $a_{\mu}(x)$ and
$\varphi(x)$, the linearized Einstein's equations are given by 
\begin{equation}
\begin{split}\label{lineinsteineqn}
R^{(1)}_{\mu\nu}=&\frac{1}{2}\partial_{\mu}\phi\partial_{\nu}\varphi+\frac{1}{2}\partial_{\mu}\varphi\partial_{\nu}\phi-\frac{V}{2}(h_{\mu\nu}- G_{\mu\nu}\delta\varphi)\\
+&\frac{Z}{2}\left[G^{\rho\sigma}F_{\mu\rho}f_{\nu\sigma}+G^{\rho\sigma}f_{\mu\rho}F_{\nu\sigma}-h^{\rho\sigma}F_{\mu\rho}F_{\nu\sigma}+\lambda\varphi G^{\rho\sigma}F_{\mu\rho}F_{\nu\sigma} \right]\\
&-Z\left[\frac{1}{4}G_{\mu\nu}(F_{\rho\sigma}f^{\rho\sigma}-g^{\rho\alpha}h^{\sigma\beta}F_{\rho\sigma}F_{\alpha\beta})+\frac{1}{8}h_{\mu\nu}F_{\rho\sigma}F^{\rho\sigma}+\frac{1}{8}\lambda\varphi G_{\mu\nu}F_{\rho\sigma}F^{\rho\sigma} \right]\ ,
\end{split}
\end{equation}
where
\begin{equation}
\begin{aligned}
R^{(1)}_{\mu \nu}&=\frac{1}{2}[\nabla_{\alpha}\nabla_{\nu}h^{\alpha}_{\mu}+\nabla_{\alpha}\nabla_{\mu}h^{\alpha}_{\nu}-\nabla_{\alpha}\nabla^{\alpha}h_{\mu \nu}-\nabla_{\nu}\nabla_{\mu}h]\ ;\\
\quad f_{\mu \nu}&= \partial_{\mu}a_{\nu}-\partial_{\nu}a_{\mu}\ ; \qquad h=G^{\mu\nu}h_{\mu\nu}\ ; \qquad \delta =\frac{2\theta/d_i}{\sqrt{2(d_i-\theta)(z-\theta/d_i-1)}}\ . 
\end{aligned}
\end{equation}
Similarly, the linearized Maxwell Equations \eqref{gaugeeqn} are
\begin{equation}\label{lingaugeeqn}
\nabla_{\mu}(Z\,f^{\mu \nu})-\nabla_{\mu}(Z\,h^{\mu \rho}F_{\rho}^{\ \ \nu})-Z(\nabla_{\mu}h^{\nu \sigma})F^{\mu}_{\ \ \sigma} +\frac{1}{2}(\nabla_{\mu}h)Z\,F^{\mu \nu}+\lambda\, Z\,F^{\mu \nu}\partial_{\mu} \varphi=0\ .
\end{equation}
The linearized scalar field equation is:
\begin{equation}
\begin{split}\label{linscalareqn}
\frac{1}{\sqrt{-G}}\partial_{\mu}(\sqrt{-G}G^{\mu \nu}\partial_{\nu}\varphi) &-\frac{1}{\sqrt{-G}}\partial_{\mu}(\sqrt{-G}h^{\mu \nu}\partial_{\nu}\phi)+\frac{1}{2}G^{\mu \nu}\partial_{\nu}\phi\partial_{\mu}h  +V\delta^{2}\varphi\\
&-\frac{\lambda Z}{4}(2F_{\mu\nu}f^{\mu\nu}-2G^{\mu\rho}h^{\nu\sigma}F_{\mu\nu}F_{\rho\sigma}+\lambda\varphi F_{\mu\nu}F^{\mu\nu})=0\ .
\end{split}
\end{equation}
In \eqref{lineinsteineqn}, \eqref{lingaugeeqn}, \eqref{linscalareqn},
indices are raised and lowered using \eqref{hvmetric}.

\section{Solution for gauge field perturbation $a_y$: Details}
\label{F2-details}

\underline{\textbf{\bm{$G_0(r)$}:}}\ \ using \eqref{F0-eqn} and \eqref{dil-g-F1}
and the variables in \eqref{xmnA-defn}, we can simplify \eqref{G0-r-eqn} to
\begin{equation}
\frac{d^2G_0}{dx^2}+\left(\frac{2m-n+1}{x}-\frac{m}{x(1-x^m)}\right)\frac{dG_0}{dx}-\frac{mn}{x^2(1-x^m)}G_0+\frac{A}{x^2(1-x^m)}=0\ .
\end{equation}
$G_0(r)$ in \eqref{G0-soln} can then be obtained as the solution to the
above equation.

\vspace{1mm}

\noindent
\underline{\textbf{$\bm{F_2(r,t,x)}$ solution:}}\ \ 
The equation governing $F_2(r)$ follows from $O(\lambda^2)$ terms of \eqref{F-eqn}. The relevant terms following from the first two lines of \eqref{F-eqn} are computationally straightforward to derive. Let us concentrate on the last line of \eqref{F-eqn}. Concentrating on the powers of $\lambda$, the integral in the last term can be integrated by parts and rewritten as
\begin{equation}
\begin{aligned}
\lambda^3 f^{\frac{i\lambda\mathbf{\Omega}}{2}}\int ds\ f^{-\frac{i\lambda\mathbf{\Omega}}{2}}\ &s^{d_i+1-z-\theta}G
\approx  \lambda^3f^{\frac{i\lambda\mathbf{\Omega}}{2}}\left[f^{-\frac{i\lambda\mathbf{\Omega}}{2}}\int ds \ s^{d_i+1-z-\theta} G_0\right.\\
&\left. +\frac{i\lambda \mathbf{\Omega}}{2}\int ds\ f^{-\frac{i\lambda\mathbf{\Omega}}{2}}\ \frac{f'}{f}\int ds'\ s'^{d_i+1-z-\theta}G_0\right]\sim  O(\lambda^3)+O(\lambda^4)\ .
\end{aligned}
\end{equation}
The above expression shows that the leading contribution from the last term of \eqref{F-eqn} becomes relevant at $O(\lambda^3)$ and has no role to play in determining $F_2(r)$. We can write the equation governing $F_2$ as
\begin{equation}
\begin{split}
F''_2-\frac{H'}{H}F'_2+\left[-i\mathbf{\Omega}\frac{f'}{f}\ F'_1-\right.&\frac{i\mathbf{\Omega}}{2}\left(\frac{f'}{f}\right)'F_1-\frac{\mathbf{\Omega}^2}{4}\left(\frac{f'}{f}\right)^2 C_0+\frac{i\mathbf{\Omega}}{2}\frac{f'}{f}\frac{H'}{H}F_1\\
&\left.+(2\pi T)^2\frac{r^{2z-2}}{f^2}\cdot(\mathbf{\Omega}^2-(2\pi T)^{2/z-2}\mathbf{Q}^2fr^{2-2z})C_0\right]=0\ .
\end{split}
\end{equation}
Using \eqref{gen-F1} and integrating the above once, we get
\begin{equation}
\begin{split}
F'_2(r)& =\frac{i\mathbf{\Omega}}{2}\frac{f'}{f}F_1-H\tilde{C}_1\\
&-\frac{C_0m^2r_0^m}{4}\left(\frac{(r_0r)^{n+2-m}}{n+2-m}\ \hyperg \left[1,\frac{n+2-m}{m},\frac{n+2}{m},(r_0r)^m\right] +\frac{1}{m}\log f\right)\cdot H
\end{split}
\end{equation}
where $\bar{C}_1$ is an arbitrary integration constant. Choosing
$\tilde{C}_1=\frac{C_0}{4}mr_0^m \left( \gamma+\psi
\left(\frac{n+2}{m}\right)\right)$ where $\gamma$ is the
Euler-Mascheroni constant and $\psi$ is the digamma function, ensures
$F'_2(r)$ is finite at the horizon (for $z<d_i+2-\theta$). The above
expression can be integrated again subject to the boundary condition
$F_2(r \sim {1 \over r_0})=0$ to obtain an explicit expression for the
function $F_2(r)$.

\vspace{1mm}

\noindent
\underline{\textbf{$\bm{G_1(r,t,x)}$ solution:}}\ \ Collecting $O(\lambda)$
terms from \eqref{G-full-eqn} gives the equation governing $G_1(r)$, 
\begin{equation}
\label{G1-r-eqn}
\begin{split}
&G''_1+\partial_r \log fr^{d_i+3-z-\theta}\ G'_1 -\frac{k^2}{r^2f}\ G_1 = \frac{i\mathbf{\Omega}}{2}\frac{f'}{f}(G'_0+\partial_r \log fr^{d_i+3-z-\theta}\ G_0) +\frac{i\mathbf{\Omega}}{2}\left(\frac{f'}{f}\right)'G_0\\
&+(2\pi T)^{\frac{1}{z}-2}\frac{\mathbf{Q}k}{r^2f}\left(\frac{C_0}{4}m^2r_0^m\left[\frac{(r_0r)^{n+2-m}}{n+2-m}\ \hyperg \Big[1,\frac{n+2-m}{m},\frac{n+2}{m};(r_0r)^m\Big] +\frac{1}{m}\log f\right] + \tilde{C}_1\right)\ 
\end{split}
\end{equation}
The homogeneous part of the above equation is identical to the homogeneous part of \eqref{G0-r-eqn}. This helps us in writing down the solution for $G_1$ as
\begin{equation}
G_1(x)=\bar{C}_1\ y_1(x) -y_1(x)\int \frac{h(x)y_2(x)}{W(x)}\ dx+y_2(x)\int \frac{h(x)y_1(x)}{W(x)}\ dx
\end{equation}
where $x$ is the radial variable defined in \eqref{xmnA-defn}. Also,
\begin{equation}
\begin{aligned}
&y_1(x)=x^{-m}\ \frac{n+(m-n)x^m}{n} ,\ y_2(x)=x^n\ \hyperg\left[2,\frac{n}{m},2+\frac{n}{m};x^m\right] ,\ W(x)=\frac{(m+n)x^{n-m-1}}{1-x^m} ,\\ & h(x)=\frac{\Lambda_1}{x^2}+\frac{\Lambda_2}{x^2(1-x^m)}+\frac{\Lambda}{x^2(1-x^m)}\left[\frac{x^{n+2-m}}{n+2-m}\ \hyperg \Big[1,\frac{n+2-m}{m},\frac{n+2}{m};x^m\Big] +\frac{\log f}{m}\right]
\end{aligned}
\end{equation}
where the constants $\Lambda_1$, $\Lambda_2$ and $\Lambda$ are given by
$\Lambda_1= -\frac{\Lambda}{m} (2-\frac{n}{m} )$,
$\Lambda_2 = \Lambda (\gamma +\psi (\frac{2+n-m}{m} ) )$,
$\Lambda=C_0kq\ r_0^{m-n-2}$.
In principle we can fix the constant $\bar{C}_1$ by demanding regularity
of the solution near horizon \ie\ as $x \sim 1$. In practice, this
appears difficult analytically.

The crucial thing to note about the $a_y$ solution is that even the
leading order piece $G_0 \sim \frac{q}{\omega}$ vanishes in the zero
momentum sector \ie\ when $q=0$. However, Kubo's formula for response
functions (the $\langle a_y(k) a_y(-k)\rangle$ correlator here) is
strictly evaluated at zero momentum: this leads to $\langle a_y(k)
a_y(-k)\rangle=0$. This is similar to the behaviour of the
correlators $\langle T^y_t(k)T^{y}_t(-k)\rangle$ and $\langle
T^y_t(k)T^{y}_x(-k)\rangle$ given by \eqref{yt-corr} which vanishes in
the $q \rightarrow 0$ limit. Indeed from
\eqref{nu-t-eqn}-\eqref{chi-eqn} we observe that when $q =0$, the
$H_{ty}$ and $a_y$ fields decouple and the shear mode sector is
governed exclusively by the dynamics of $H_{xy}$. This results in a
non-trivial $\langle T^y_x(k)T^y_x(-k)\rangle$ correlator
\ie\ \eqref{xy-xy-corr} which is indeed non-zero in the $q=0$ sector
eventually leading to \eqref{kubo}.

\section{Shear diffusion and ${\eta \over s}$ from membrane paradigm}

Here we briefly review \cite{Kolekar:2016pnr,Kolekar:2016yzg} which
motivated the present work. The analysis there adapted the study in 
\cite{Kovtun:2003wp} of obtaining a diffusion equation from near
horizon shear gravitational perturbations in a membrane paradigm type
approach to hvLiv theories (\ref{hvmetric}).

The simplest context is dilaton gravity with $A_{\mu}=0$ in
\eqref{hvaction} which fixes $z=1$ from (\ref{bg-gauge-fld}).  The
spatial directions $\{x_i\}$ enjoy translation invariance. Thus the
diffusion of shear gravitational modes $h_{xy}$, $h_{ty}$, can be
mapped to charge diffusion in an auxilliary theory obtained by
compactifying the $(d_i+2)$-dimn theory along one of the spatial
directions, say $y$.
We turn on plane wave modes for the perturbations
$\propto e^{-\Gamma t+iqx}$ where $\Gamma$ is the typical time scale
over which the perturbation decays while $q$ is the momentum along
$x$.  The $y$-compactification maps $h_{xy}, h_{ty}$ to gauge fields
$\scriptA_t=g^{-1}_{yy}h_{ty}$ and $\scriptA_x=g^{-1}_{yy}h_{xy}$
(identical to the field variables in \eqref{new-var}) in the 
$d_i+1$-dimensional theory. These satisfy the Maxwell Equations
\ie\ $\partial_{\mu}(\frac{1}{g_{eff}^2}\sqrt{-g}\scriptF^{\mu
  \nu})=0$. The field strength is defined as usual as
$\scriptF_{\mu\nu}=\partial_{\mu}\scriptA_{\nu}-\partial_{\nu}\scriptA_{\mu}$
while the $r$-dependent coupling is\
$1/g^2_{eff}=(g_{yy})^{\frac{d_i}{d_i-1}}$. Then defining currents
$j^{\mu}=n_{\nu}\scriptF^{\mu\nu}\equiv (j^t,j^x)$ on the stretched
horizon, with $n$ the outward unit normal, a set of approximations
allows using these equations to eventually obtain Fick's Law\
$j^x=-\mathcal{D}\partial_xj^t$ for charge diffusion of the gauge field
$\scriptA_{\mu}$ in the compactified background: we then identify the
diffusion constant.

In hvLif theories with a gauge field, the metric perturbations couple
to the gauge field perturbations which complicates formulating Fick's
Law. However, we note that these are uncharged black branes with the
gauge field and scalar simply serving as sources supporting the
nonrelativistic background. Using intuition from the fluid/gravity
correspondence \cite{Bhattacharyya:2008jc}, we expect that the near
horizon perturbations must be characterized simply by local
temperature and velocity fluctuations. Thus since charge cannot be a
parameter, we expect that that the structure of the diffusion equation
and the shear diffusion constant should not be dramatically altered by
the presence of the gauge field. In this light, we see that
\eqref{nu-r-eqn} with explicit $t$ and $x$ derivatives is schematically
of the form\ $\partial_x(\#\partial_rh_{xy}) \sim \partial_t (\# \partial_r
h_{ty}-\# a_y)$, using \eqref{nu-r-eqn}. This suggests the use of new
field variables $\tilde{h}_{xy}$ and $\tilde{h}_{xy}$,
\begin{equation}\label{tildeh}
  \tilde{h}_{ty}\equiv h_{ty}-kr^{2\theta/d_i-2}\int_{r_c}^r s^{d_i+1-z-\theta} a_y\  ds\ ,
  \qquad \quad \tilde{h}_{xy} \equiv h_{xy}\ ,\\
\end{equation}
or equivalently $\tilde{\scriptA}_t$ and $\tilde{\scriptA}_x$ in the
compactified theory. This is consistent with and motivated \eqref{E-defn}.
The field $\chi$ is an effective scalar field arising from the
compactification of $a_y$.
Defining the currents $\tilde{j}^x=j^x=n_r\scriptF^{xr}$ and
$\tilde{j}^t=n_r\tilde{\scriptF}^{tr}=-n_r\partial^r
\tilde{\scriptA}_t$ on the stretched horizon we can express Fick's Law
as $\tilde{j}^x=-\mathcal{D}\partial_x \tilde{j}^t$, which we outline below.

In the near horizon region the equations of motion simplify significantly
if we make the following assumptions: (a) $\tilde{\scriptA}_{\mu}$ admits a
series expansion in the parameter $\frac{q^2}{T^{2/z}}$ \ie\ 
\begin{equation}
\tilde{\scriptA}_{\mu} (t,x,r)=\tilde{\scriptA}_{\mu}^{(0)}(t,x,r)+\frac{q^2}{T^{2/z}}\tilde{\scriptA}^{(1)}_{\mu}(t,x,r)+O\left(\frac{q^4}{T^{4/z}}\right)+\cdots \qquad \mu \equiv (t,x)\ ,
\end{equation}
and also, (b) $|\partial_t \tilde{\scriptA}_x| \ll |\partial_x \tilde{\scriptA}_t|$. The regime of validity of this analysis is
\begin{equation}
\label{rh-bound}
e^{-\frac{T^{2/z}}{q^2}} \ll \frac{\frac{1}{r_0}-r_h}{\frac{1}{r_0}} \ll \frac{q^2}{T^{2/z}} \ll 1\ .
\end{equation}

The field strength $\tilde{\scriptF}_{tx}$ follows a wave equation: we 
then choose the ingoing solution. Equivalently, this condition is
obtained by demanding that the function $H\equiv r^{\#}fH'_{xy}$ depends
only on the ``infalling'' coordinate $v \sim t+\log (1/r_0-r)$, which gives
\begin{equation}
\tilde{\scriptF}_{tx}+(d_i+z-\theta)r_0^z\left(\frac{1}{r_0}-r\right)\tilde{\scriptF}_{rx}=0\ .
\end{equation}
Alongwith a series of approximations and simplications, we eventually 
obtain Fick's Law $\tilde{j}^x=-\mathcal{D}\partial_x \tilde{j}^t$
for the variables (\ref{tildeh}) on the stretched horizon.

For $z<d_i+2-\theta$, the leading order solution for the gauge field
$\tilde{\scriptA}_t$ and the shear diffusion constant $\mathcal{D}$ are
\begin{equation}
  \tilde{\scriptA}_t^{(0)}=\frac{Ce^{-\Gamma t+iqx}}{d_i+2-z-\theta}\
  r^{d_i+2-z-\theta}\ , \qquad
\mathcal{D}=\frac{r_0^{z-2}}{d_i+2-z-\theta}\ .
\end{equation}
The expression for $\tilde{\scriptA}_x^{(0)}$ can be found from
$\tilde{\scriptA}_t^{(0)}$.
Using the above expression for $\mathcal{D}$ along with \eqref{tempr0}
leads to \eqref{genDiffconst}. In \cite{Kolekar:2016pnr,Kolekar:2016yzg},
we then conjectured \eqref{eta-s-D-T} by examining various special cases: (i)
For pure $AdS$ when $z=1, \theta=0$, we obtain $\mathcal{D}=\frac{1}{4\pi T}$
which along with the thermodynamic relation ${\eta \over
  s}=\mathcal{D}T$ leads to ${\eta \over s}={1 \over 4\pi}$;\ (ii)
From \eqref{genDiffconst} it follows for any hyperscaling violating
theory with $z=1$, the $\theta$-dependent prefactors cancel precisely
giving ${\eta\over s}={1 \over 4\pi}$~. This is vindicated for
$z=1, \theta \neq 0$ hvLif theories arising from the reduction of
non-conformal $Dp$-branes on the sphere $S^{8-p}$ \cite{Dong:2012se}:
it is well-known that the latter satisfy the viscosity bound and we
expect that long-wavelength physics is unaffected by the sphere
reduction;\ (iii) For theories with exact Lifshitz scaling symmetry
\ie\ $z\neq 0$ and $\theta =0$, the diffusion equation implies
the scaling dimension $dim [\mathcal{D}] \sim z-2$, where $[x_i]=-1$
and $[t]=-z$. Earlier investigations on Lifshitz hydrodynamics
\cite{Hoyos:2013qna,Kiritsis:2015doa} argued that ${\eta \over s}$
exhibits universal viscosity behaviour. This again vindicates the
proposed relation \eqref{eta-s-D-T} for consistency.  This leads us to
the universal viscosity bound (\ref{eta-s-D-T}) for $z<d_i+2-\theta$,
the sector continuously connected to $AdS$.

When $z=d_i+2-\theta$, the leading solution as well as the shear
diffusion constant scale logarithmically,
\begin{equation}
  \tilde{\scriptA}_t^{(0)}=Ce^{-\Gamma t+iqx}\log \frac{r}{r_c}\ ,\qquad
  \mathcal{D}=r_0^{z-2}\log \frac{1}{r_0r_c}\ .
\end{equation}
This is reflected in the quasinormal modes through \eqref{qnmHlogcase},
\eqref{logdiff}. It is likely that this leading shear diffusion constant
undergoes some renormalization/resummation. We note also that
$\frac{\eta}{s}$ continues to exhibit universal viscosity behaviour
in the Kubo limit as we have seen in sec.~3.1: perhaps some aspects
of hydrodynamics are novel in this case.
									
For $z>d_i+2-\theta$, the series solution breaks down in the near
horizon region: the boundary conditions are not satisfied suggesting
non-universal UV physics dominates, and we obtain no insight on the
shear diffusion constant.

\end{document}